\documentclass{ws-ijmpcs}

\begin{document}

\markboth{P. Banasi\'nski \& W. Bednarek}
{THE OPTICALLY THICK HOMOGENEOUS SSC MODEL: APPLICATION TO RADIO GALAXY NGC 1275}

%
\catchline{}{}{}{}{}
%

\title{THE OPTICALLY THICK HOMOGENEOUS SSC MODEL: APPLICATION TO RADIO GALAXY NGC 1275}

\author{PIOTR BANASI\'NSKI and WLODEK BEDNAREK}

\address{Department of Astrophysics, University of Lodz,\\
ul. Pomorska 149/153, 90-236 Lodz, Poland\\
pgbanasinski@gmail.com}

\maketitle

\begin{history}
\received{Day Month Year}
\revised{Day Month Year}
\end{history}

\begin{abstract}
We consider the Synchrotron Self-Compton (SSC) model for jets in active galaxies in which produced $\gamma$-ray photons can be  
absorbed in collisions with the synchrotron radiation already at the emission region. In terms of such modified SSC model, we argue that the higher emission stages should be characterised by $\gamma$-ray spectra extending to lower energies due to the efficient absorption of the highest energy $\gamma$-rays. As an example, we show that different emission stages of the nearby radio galaxy NGC 1275
could be explained by such scenario.
\end{abstract}

\ccode{PACS numbers: 98.54.Gr; 98.62.Js; 98.58.Fd; 95.85.Pw}

\section{Introduction}

The $\gamma$-ray emission from active galaxies is popularly interpreted in terms of homogeneous Synchrotron Self-Compton model in which single population of electrons, distributed homogeneously in specific (usually spherical) volume, produce synchrotron radiation which is up-scattered to $\gamma$-ray energies in the Inverse Compton (IC) process.  Usually, the parameters of the emission region are chosen in such a way that produced $\gamma$-rays escape without significant absorption. 
In reality this does not need to be always the case. For emission regions compact enough, produced $\gamma$-rays can be absorbed in this same synchrotron radiation field (see e.g. Sitarek \& Bednarek 2007). In this paper we consider the SSC models with the parameters for which the absorption of $\gamma$-rays can not be neglected. Such optically thick SSC models predict specific emission features
from the active galaxies in different emission stages which could be searched for GeV-TeV $\gamma$-ray observations.

\section{Optically thick SSC model}

We consider the standard homogeneous SSC model for the specific emission region in the jet of 
active galaxy. The emission region (a blob) is assumed to be spherical with the radius $R$. 
It moves along the jet with the specific Doppler factor $D$. The radius of the emission region can be constrained from the observed variability time scale of emission, $\tau_{\rm v}$, and the applied Doppler factor of the blob, $R\approx 0.5cD\tau_{\rm v}$. Relativistic electrons, injected into the blob,
produce synchrotron radiation in the magnetic field $B$. These synchrotron photons are comptonized by these same electrons to TeV $\gamma$-ray energies. For the blobs compact enough, produced $\gamma$-rays can be additionally absorbed in the blob synchrotron radiation. Based on the observed synchrotron spectrum, we derive the electron spectrum for assumed value of the magnetic field $B$. This electron spectrum allows us to calculate the GeV-TeV $\gamma$-ray spectrum produced in the SSC radiation process.
Knowing the synchrotron radiation field within the blob, we calculate the absorption effects on IC $\gamma$-rays during their propagation through the blob. In order to do that we calculate the optical depths for $\gamma$-rays with specific energies. As an example, we apply the synchrotron spectra within the blob corresponding to the emission stages observed from the radio galaxy NGC 1275.
The results of calculations, for different emission stages of NGC 1275 (Fig.~1 on the right)  and specific parameters of the emission region ($\tau_{\rm v}$, $D$), are shown on the left Fig.~1. 
In fact, the radio emission shown in Fig.~1 may not come from the blob region. However, this low energy photons do not contribute to the optical depth of $\gamma$-rays with energies reported in Fig.~1 (on the left).
Note that for selected parameters, the optical depths are close to unity at energies of $\gamma$-ray photons $\sim$100 GeV. We conclude that in such case, the absorption of $\gamma$-rays should have important effect on the shape of the $\gamma$-ray spectrum in the range of the emission stages observed from NGC 1275.

\begin{figure}
\vskip 6.truecm
\includegraphics{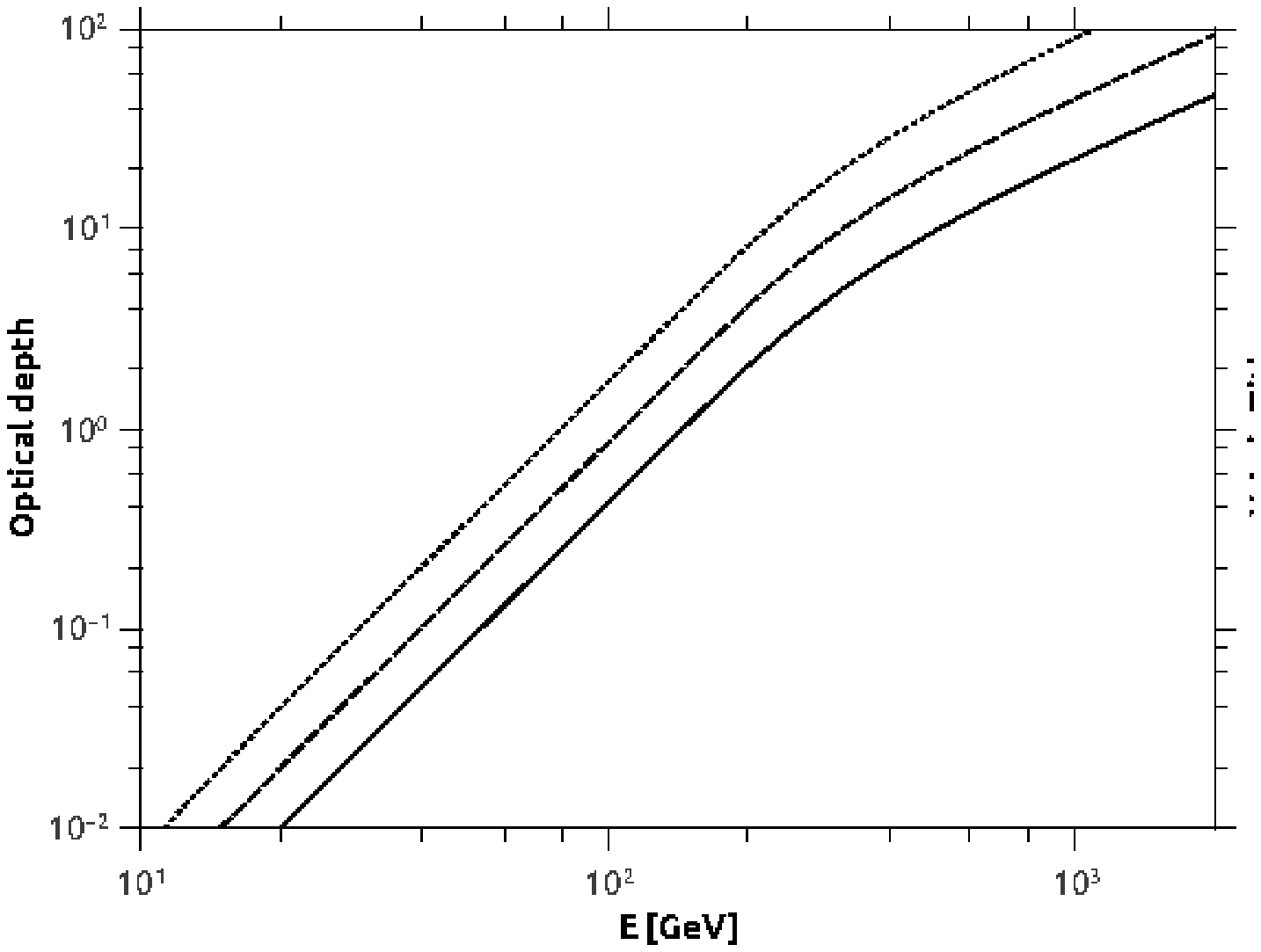}
\includegraphics{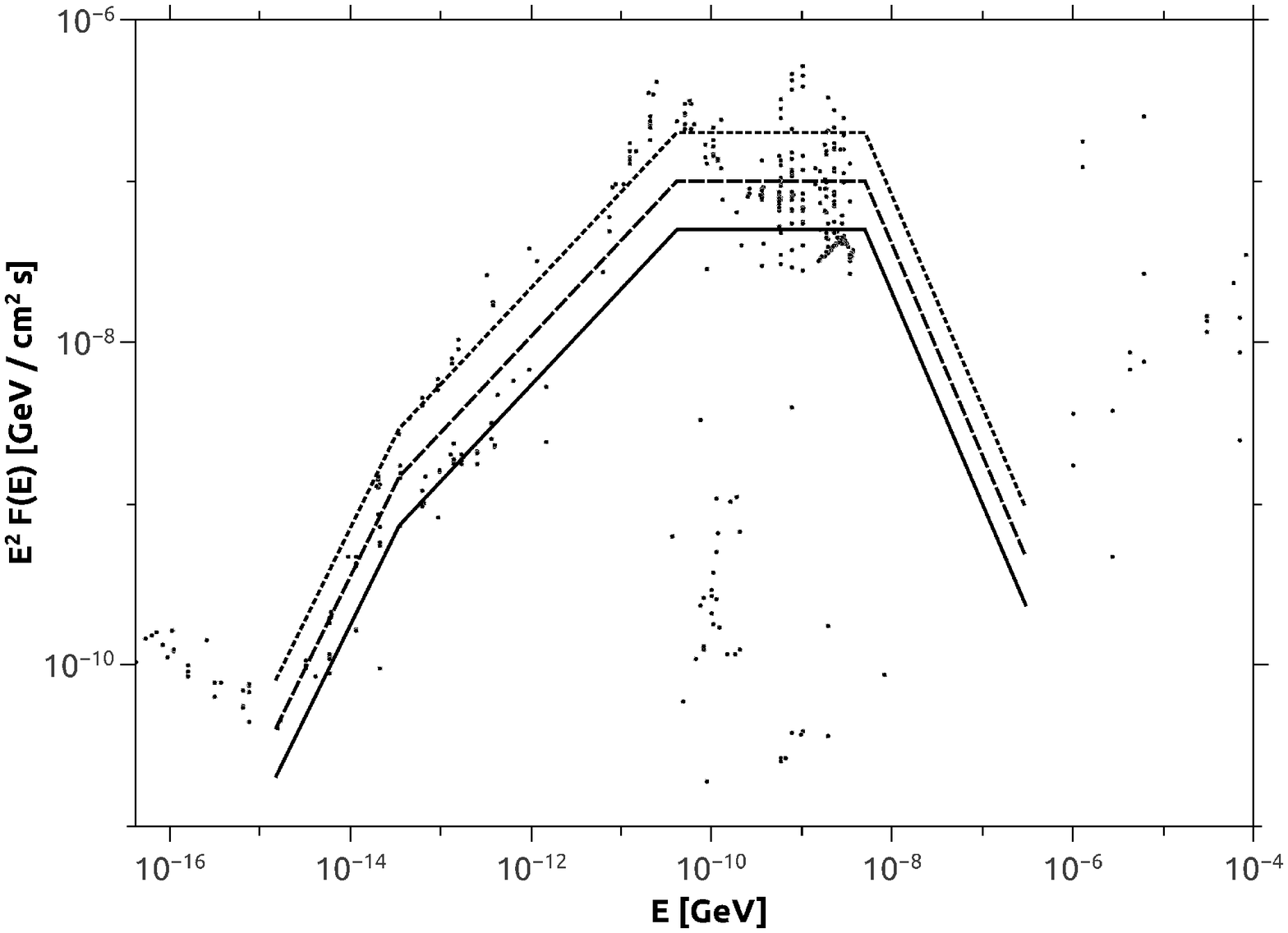}
\caption{Example optical depths for $\gamma$-rays in the synchrotron radiation of the blob. $\gamma$-rays and synchrotron radiation is  produced by relativistic electrons. We show the optical depths for fixed parameters of the blob (radius $R = 9\times 10^{15}$ cm, Doppler factor $D = 2$) but for different density of electrons in the blob. Different curves correspond to different levels of synchrotron emission. The spectra of electrons have been derived from 
the observations of the synchrotron spectrum of NGC 1275 (see NED) as shown on the right.
}
\label{fig1}
\end{figure}
\section{Application of the model to radio galaxy NGC 1275}

As an example, we apply optically thick SSC model for different emission stages observed in the 
radio Galaxy NGC 1275. This galaxy has been detected by the Fermi-LAT telescope showing the power law spectrum with the spectral index not far from $-2$ (Abdo et al.~2009). The GeV emission is strongly variable on month time scales (Kataoka et al. 2010) and a few days time scale (Brown \& Adams~2011) showing the flux variability by a factor of a few. NGC 1275 has been also detected at TeV energies  having the steep spectrum between 70-500 GeV (Aleksi\^c et al.~2012). The spectral index in this energy range is $-4.1$, suggesting the presence of a break or cut-off around tens of GeV. The sub-TeV emission shows only weak hint of variability in this energy range on a month time scale (Aleksi\^c et al.~2013).

Such spectral behaviour can be naturally interpreted in terms of the optically thick SSC model.
As we have shown above, this model predicts that higher emission stage in GeV energies requires more electrons injected into the blob. These larger number of electrons should also produce stronger 
synchrotron radiation which is responsible for stronger absorption of TeV $\gamma$-rays. As a result, 
GeV emission from active galaxy can vary strongly, but the corresponding TeV emission does not change
significantly.  In order to confirm such behaviour, we perform calculations of the $\gamma$-ray spectra in terms of optically thick SSC model applying likely range of parameters of the emission region.
We consider the variability time scale of this emission of the order of days $\tau_{\rm v} = 3\times 10^5$ s and the Doppler factor of the blob $D =2$. 
Rather low Doppler factor of the blob in NGC 1275 is due to relatively large jet viewing angle estimated on $30^\circ-55^\circ$ (Vermeulen et al. 1994, Walker et al. 1994).
These parameters allow us to estimate the radius of the blob on $R = 9\times 10^{15}$ cm. We also fix the magnetic field strength in the emission region on $B = 0.3$ G. For these parameters we calculate the $\gamma$-ray spectra escaping from the blob for two synchrotron emission stages as shown in Fig.~1 (on the right, see NED\footnote{NED: http://ned.ipac.caltech.edu/}). The results of calculations are shown in Fig.~2 together with the levels of emission derived in the mentioned above publications. Our model is consistent with the main feature of $\gamma$--ray emission from NGC 1275. Although, the GeV flux is strongly variable, we obtained that the TeV $\gamma$-ray flux should stay approximately constant during different GeV emission stages. This effect is due to stronger absorption of sub-TeV $\gamma$-ray emission in the enhanced synchrotron radiation of the emission region.

\begin{figure}[t]
\centerline{\psfig{file=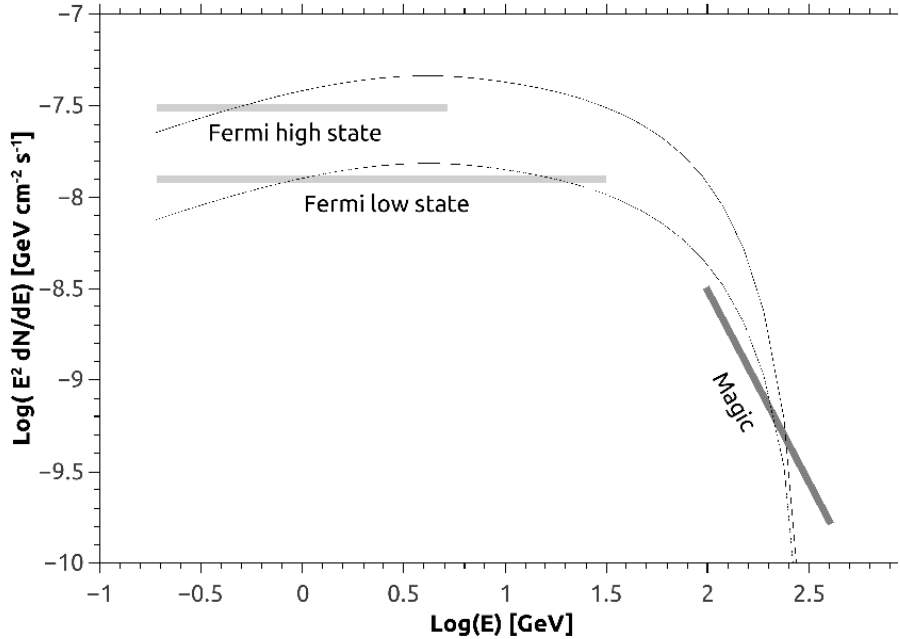,width=12.cm}}
\vspace*{10pt}
\caption{Comparison of the example calculations in terms of the optically thick SSC model with emission stages of radio galaxy NGC 1275 in the GeV $\gamma$-ray energy range (thick light line - FERMI data, Aleksi\^c et al.~2013), and TeV energies (thick dark line - MAGIC, Aleksi\^c et al.~2013). The emission stages are defined by different levels of GeV emission (and also synchrotron emission) which are characterised by different opacity of the emission region. The parameters of the emission region are the following: radius of the blob $R = 9\times 10^{15}$ cm, $\tau_{\rm v} = 3\times 10^5$ s, and magnetic field strength $B = 0.3$ G. \label{f1}}
\end{figure}

\section{Conclusion}

We analyse the emission features of the optically thick homogeneous SSC model. It is shown that such model predicts relatively weak variability of TeV $\gamma$-ray emission from some active 
galaxies in which case the GeV $\gamma$-ray emission is strongly variable. Such spectral behaviour is due to strong absorption of sub-TeV $\gamma$-ray photons by synchrotron radiation already at the blob.
We show that this model is consistent with the $\gamma$-ray emission features recently reported from the radio galaxy NGC 1275.

\section*{Acknowledgements}
This work is supported by the grant through the Polish Narodowe Centrum Nauki No. 2011/01/B/ST9/00411.



\begin{thebibliography}{0}    

\bibitem{ab10} A.A. Abdo, et al. {\it ApJ} {\bf 699} 31 (2009).
\bibitem{aletal12} J. Aleksi\^c et al. {\it A\&A} {\bf 539} L2 (2012).
\bibitem{aletal13} J. Aleksi\^c et al. {\it A\&A}, submitted (2013). 
\bibitem{ba11} A.M. Brown, J. Adams {\it MNRAS} {\bf 413} 2785 (2011).
\bibitem{sb07} J. Sitarek, W. Bednarek, {\it ApSS} {\bf 309} 105 (2007).
\bibitem{ketal10} J. Kataoka et al {\it ApJ} {\bf 715} 554 (2010).
\bibitem{vetal94} R.C. Vermeulen et al. {\it ApJ} {\bf 430} L41 (1994).
\bibitem{wetal94} R.C. et al. {\it ApJ} {\bf 430} L45 (1994).

\end{thebibliography}
\end{document}